\documentclass[10pt]{article}
\usepackage{graphicx}
\newcommand{\beq} {\begin{equation}}
\newcommand{\eeq} {\end{equation}}
\newcommand{\bsigma}{\mbox{\boldmath $\sigma$}}
\newcommand{\btau}{\mbox{\boldmath $\tau$}}
\newcommand{\bqu}{{\bf q}}
\begin{document}
\begin{center}
{\Large \bf Short-range correlations in finite nuclear systems} \\
\vspace{1.0 cm}
{\Large  Giampaolo Co' } \\ 
\vspace{1.cm}
{\large Dipartimento di Fisica,  Universit\`a di Lecce, \\
and } \\
{\large Istituto Nazionale di Fisica Nucleare  sez. di Lecce, 
\\ I-73100 Lecce, Italy} 
\end{center}
\vskip 1.0 cm 
\begin{abstract}
  Recent results concerning the use of the Correlated Basis Function
  to investigate the ground state properties of medium-heavy doubly
  magic nuclei with microscopic interactions are presented.  The
  calculations have been done by considering a Short-Range Correlation
  between nucleons. The possibility of identifying effects produced by
  Short-Range Correlations in electromagnetically induced phenomena is
  discussed.
\end{abstract} 
\section{Introduction}
\label{se:intr}
In many-body physics the word {\bf correlation} indicates effects
beyond Mean-Field (MF) theories. In nuclear physics it is common to
distinguish between short- and long-range correlations.  Nuclear
collective phenomena such as vibrations and rotations are ruled by
long-range correlations. These effects are well known and studied
since the infancy of nuclear physics. On the contrary the study of the
Short-Range Correlations (SRC) is a relatively new issue in nuclear
physics. These correlations are produced by the strong repulsive core
of the microscopic nucleon-nucleon interaction at short internucleon
distances. In spite of the fact that all the microscopic nuclear
theories need the SRC, clear signatures of their presence in
medium-heavy nuclei have not yet been identified.  The search for
nuclear phenomena showing SRC effects is one of the most discussed
topics of the last 15 years in the nuclear structure community.

In this contribution I shall be concerned about SRC. In the first part
I shall present some results of microscopic calculations regarding
ground state properties of medium-heavy nuclei. These calculations
have been done with realistic nucleon-nucleon interactions and the SRC
have been used to tame the repulsive core of these interactions. In
the second part of the contribution I shall discuss the results of a
study aimed to identify the presence of SRC in nuclei by using
experiments done with electromagnetic probes.

\section{The nuclear ground state}
\label{se:ngs}
The results presented in this section, have been obtained within the
framework of the Correlated Basis Function (CBF) theory \cite{cla79}.
The starting point of this theory is the Ritz's variational
principle:
\beq
\label{eq:var}
\delta E[\Psi] \equiv 
\delta \left[ \frac  {<\Psi|H|\Psi>} {<\Psi|\Psi>} \right] = 0
\eeq
with the following ansatz for the many-body nuclear state:
\beq
\label{eq:psi}
|\Psi> = F |\Phi_0> 
\eeq
where $|\Phi>$ indicates a ground state Slater determinant constructed
with a set of orthonormal single particle wavefuctions $|\phi>$, and
$F$ is a many-body correlation function defined as:
\beq
\label{eq:corr}
F= {\cal S} \prod_{i<j} \left( \sum_{p=1}^n f^p(r_{ij}) O^p(i,j) \right) 
\eeq
In the above equation ${\cal S}$ indicates a symmetrizer
operator. 
The two-body correlation functions $f^p(r_{ij})$ have an
operatorial dependence of the same type of that of the nucleon-nucleon
interaction. In the present context I shall consider correlations up
to $p=8$ of the type:
\beq
\label{eq:operators}
O^{p=1,8}(i,j) = 
[ 1, \, \bsigma(i) \cdot \bsigma(j), \,S(i,j), 
{\bf L}(i,j) \cdot {\bf S}(i,j)] 
\otimes [\btau(i) \cdot \btau(j)]
\eeq
where $S(i,j)$ is the usual tensor operator.

The problem is now well defined, and the minimization of the energy
functional (\ref{eq:var}) is, at least in principle, only a technical
matter. Monte Carlo techniques have been used \cite{pud97} to solve
the multidimensional integrals of eq. (\ref{eq:var}).  This approach
is {\sl exact}, i.e. no further approximations are required, but it
can be used only for relatively light nuclei. So far it has been
applied to nuclei up to A=8, and for the $^{16}$O nucleus the Monte
Carlo calculations have been done with an approximation consisting in
considering only up to four-body nucleon clusters \cite{pie92}.

It is clear that the minimization of the energy functional
(\ref{eq:var}) for heavier nuclei requires the use of techniques
alternatives to the Monte Carlo. These new techniques should simplify
the problem by including new approximations. The idea is to find
approximations affecting only irrelevant parts of the calculation, in
order to obtain accurate results.

A technique rather well studied is the Fermi Hypernetted Chain (FHNC)
theory \cite{cla79}, developed first for infinite systems such as
liquid helium and nuclear matter. A first step in the application of
the FHNC consists in a topological analysis of the various nucleon
clusters formed by the interaction, by the SRC and by the
antisymmetrization of the many-body wave function. It is possible to
sum in a closed form all the diagrams with certain topological
characteristics; these are called {\bf nodal} diagrams. The other
diagrams, named {\bf elementary}, have to be calculated one by one, as
in traditional perturbation theory. In this approach the most common
approximation consists in neglecting the elementary diagrams. This
approximation, called FHNC/0, is rather good in nuclear matter, even
though it is not accurate enough for helium systems since they
have a relatively higher density.

The FHNC theory has been extended to finite nuclear systems in refs.
\cite{fan79}, and applied to light doubly magic nuclei in refs.
\cite{co92,ari96} for simple interactions and correlations.  The main
goal of that investigation was a test of the validity of the theory.
We found that the various sum rules were exhausted at the 0.1\% level
when a specific elementary diagram was included.  We named this
approximation FHNC-1.
\vskip 0.5 cm
\begin{table} [h]
\begin{center}
\begin{tabular}{|ccc|ccc|}
\hline
\multicolumn{3}{|c|}{$^{16}$O} &
\multicolumn{3}{|c|}{$^{40}$Ca} \\
\hline
   full   & corr   & exp   & full    & corr & exp \\
 -5.48    & -5.41  & -7.97  & -6.97  & -6.64 & -8.55 \\
\hline
\end{tabular}
\end{center}
\caption{ Binding energies per nucleon in MeV. The label {\sl full}
 indicates that the minimization has been done on both correlation and
 single particle wave functions, while {\sl corr} indicates that only
 the correlation has been varied.
}
\label{tab:ene}
\end{table}

The next step of the work consisted in extending the theory in order
to use realistic microscopic interactions with their full operatorial
dependence. This requires to activate the analogous operatorial
dependence of the correlation function, as given in eq.
(\ref{eq:corr}).  The various terms of the correlation do not commute
any more with the hamiltonian, and among themselves. The various
cluster terms cannot be resummed in a closed form 

For this reason it is necessary to introduce in the FHNC-1
computational scheme another approximation called Single Operator
Chain. The FHNC-1 calculations in this approximation sum, in addition
to all the nodal diagrams containing the scalar term of the
correlation, only those diagrams where a single state dependent
operator terms appears.  This further approximation is slightly
worsening the accuracy of the result.  The sum rules are now exhausted
at about the 5\% level \cite{fab98}.
 
With this approach we have evaluated the ground state properties of
the $^{16}$O and $^{40}$Ca nuclei using the realistic 2-body
interaction Argonne V8' plus the 3-body Urbana IV force
\cite{fab00,fab01}.  The first one is a reduced version of the Argonne
V18 interaction limited to the first 8 channels of eq.
(\ref{eq:operators}). In order to compensate for the truncation of 6
operatorial channel the parameters of the three-body Urbana IV force
have been slightly readjusted \cite{pud97}.

The minimization procedure implied by eq. (\ref{eq:var}) involves both
correlation and single particle wave functions. The energies obtained
by this complete minimization are compared in tab. \ref{tab:ene} with
the empirical binding energies. The discrepancy may seem rather large,
but it is of the same magnitude of that of the best microscopic nuclear
matter calculations.
\begin{figure}
\vspace{-1.0 cm}
\hspace{0.5 cm}  
\includegraphics[angle=90,scale=0.6]
       {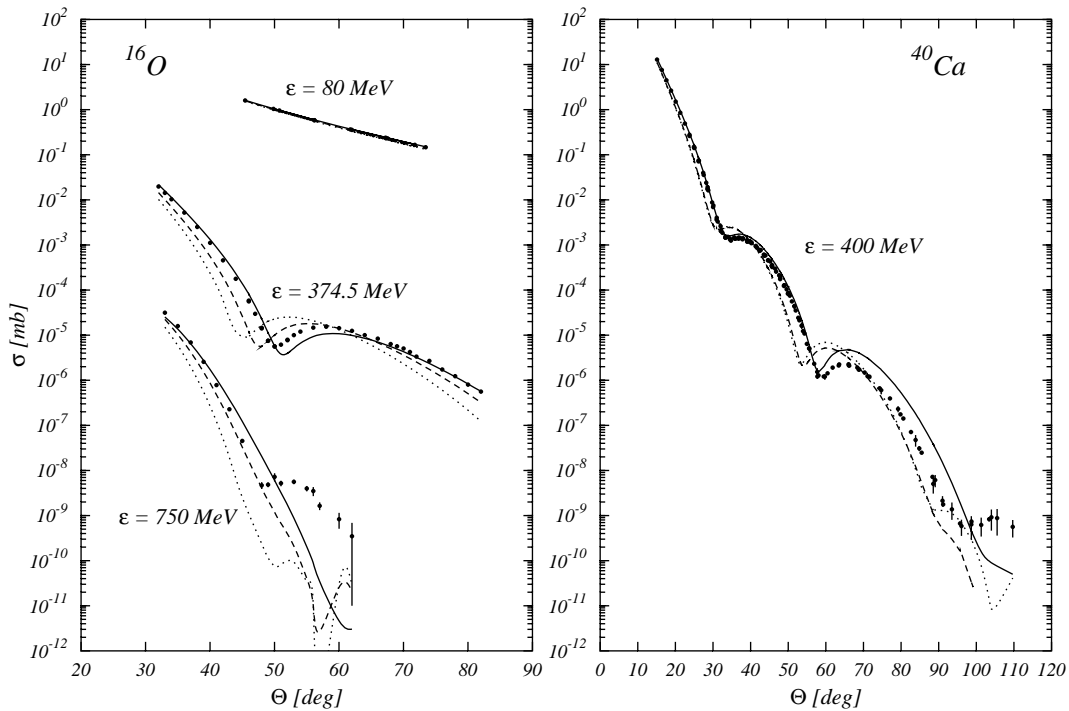}
\caption{\small Elastic electron scattering cross sections. The full
lines have been obtained with the empirical mean field wave functions
and the V8' interaction. The dashed lines have been obtained by using
the fully minimized densities with the V8' interaction and the dotted
line with the Argonne V14 interaction.
}
\label{fig:ee}
\end{figure}

The calculated charge densities distributions of $^{16}$O and
$^{40}$Ca have been used to evaluate the elastic electron scattering
cross section \cite{fab00}. These cross sections are shown by the
dashed lines of fig. \ref{fig:ee} where they are compared with the
data and the results of an analogous calculation done with the Argonne
V14 interaction (dotted lines). Also in this case the agreement with
the data is rather poor. Theory and experiments start to differ
already at 30$^0$.

The same computational scheme has been used by keeping fixed the set
of single particle wave functions.  We used a set already utilized in
ref. \cite{ari96} and chosen to reproduce charge radii and single
particle energies around the Fermi surface. The minimum of the
energy functional (\ref{eq:var}) is obtained by changing only the
correlation. The results of this calculation are shown in tab.
\ref{tab:ene} and in fig. \ref{fig:ee} by the full lines.

The difference between the binding energies obtained with a full
minimization and with the one restricted to the correlation function
only, is of the order of the few percent. This seems to indicate the
scarce sensitivity of this quantity to the single particle wave
functions. On the other hand the charge densities, and consequently
the cross sections, have been strongly modified. Now the agreement
with the data extends up to about 60$^0$.
\begin{figure}
\hspace {3.4 cm}
\includegraphics[angle=270,scale=0.5]
       {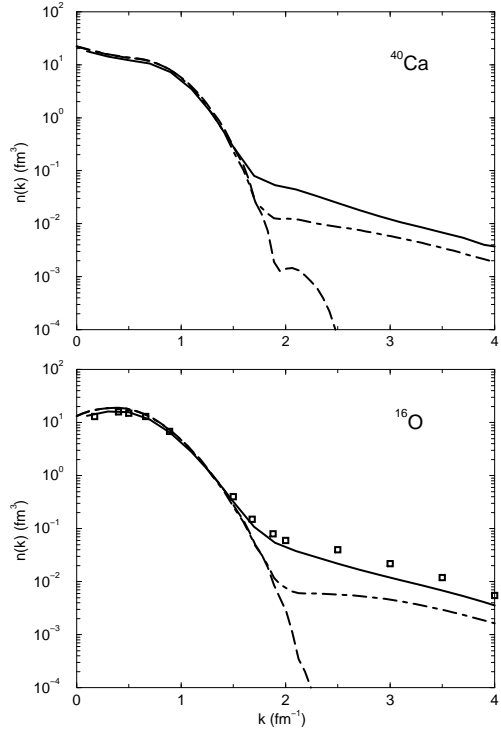}
\caption{\small Momentum distribution. The full lines have been
calculated considering central and tensor terms of the SRC. The dashed
dotted lines have been calculated with only the central SRC and the
dashed lines show the pure IPM results. The white dots are the Monte
Carlo results of ref \protect\cite{pie92}. 
}
\label{fig:nk}
\end{figure}

This last set of single particle wave functions has been used to
calculate the momentum distribution of the two nuclei considered
\cite{fab01}. The result of this calculation is shown in fig.
\ref{fig:nk}. The dashed lines show the MF distributions, the dashed
dotted lines have been obtained with purely scalar correlations and
the full lines show the results of the complete calculation.
These results show that the contribution of the state dependent terms
of the correlation is not negligible. The agreement with the Cluster
Monte Carlo momentum distribution of $^{16}$O (white squares) is
rather good. 

The large increase of the correlated momentum distributions with
respect to the MF results is a well known feature produced by the SRC
\cite{ant88}. Unfortunately the nucleon momentum distribution is not
directly observable. It is therefore necessary to find observables
related to it in order to identify the presence of SRC. This search
should be done by studying the nuclear excited states.  The results of
this investigation are the subject of the next section.

\begin{figure}
\vspace{-0.8 cm}
\hspace{4.0 cm}
\includegraphics[angle=0,scale=0.4]
       {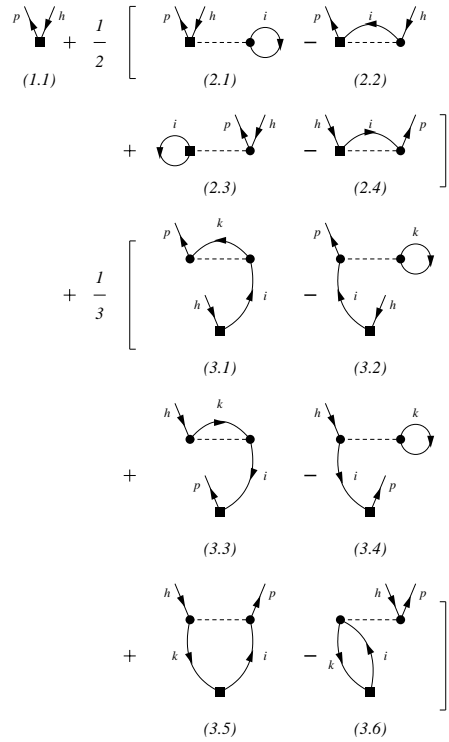}
\caption{\small Mayer-like diagrams showing the two- and tree-body
  clusters terms included in the present work. The squares indicate
  the coordinate where the one-body electromagnetic operator acting
  on.
}
\label{fig:1p1h}
\end{figure}
\section{Search for SRC effects in excitation processes}
\label{se:search}
In the framework of the CBF theory, the description of the 
excited states of the nucleus can be done by extending the ansatz of
eq. (\ref{eq:psi}):
\beq
\label{eq:psi_n}
|\Psi_n> = F |\Phi_n> 
\eeq
where now $|\Phi_n>$ indicates a Slater determinant containing
particle-hole excitations. In principle, in the description of the
excited states also the correlation function should change, and the
variational principle (\ref{eq:var}) should be again applied.
\begin{figure} [h]
\hspace{4.0 cm}
\includegraphics[angle=0,scale=0.6]
       {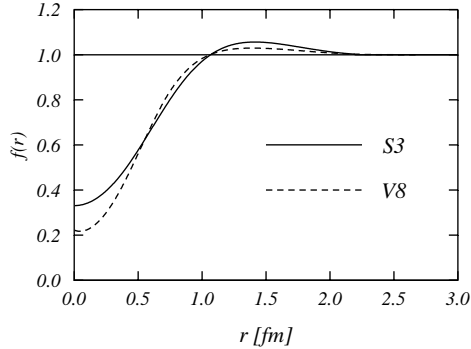}
\vspace{-0.3 cm}
\caption{\small Two-body 
  correlation functions used in the calculations of the
  electromagnetic response functions.
}
\label{fig:corr}
\end{figure}

In reality, we have used the approach of Fantoni and Pandharipande
\cite{fan87} which considers both correlation function and MF
potential fixed by the ground state minimization.  The transition
amplitude from the ground to an excited state, induced by an external
operator, can be evaluated using cluster expansion techniques
analogous to those used for the ground state. In the special case of
infinite nuclear systems, it has been shown \cite{fan87,fab89} that it
is possible to sum all the diagrams of a certain type.

Our calculations of the electromagnetic responses of finite nuclei,
have an additional approximation, consisting in retaining only the
diagrams that contain a single correlation line. This amount to consider
only certain classes of two-body and three-body cluster diagrams.  As
example I show in fig. \ref{fig:1p1h} the Mayer-like diagrams
considered in the case of an excited state which asymptotically can be
described in terms of one-particle one-hole excitation. In ref.
\cite{ama98} this approximation has been tested against the complete
calculation for the nuclear matter charge responses.  The agreement
between the two approaches is excellent. The differences are of the
order of 10$^{-5}$.
\begin{figure} [h]
\hspace{1.0 cm}
\includegraphics[angle=90,scale=0.5]
       {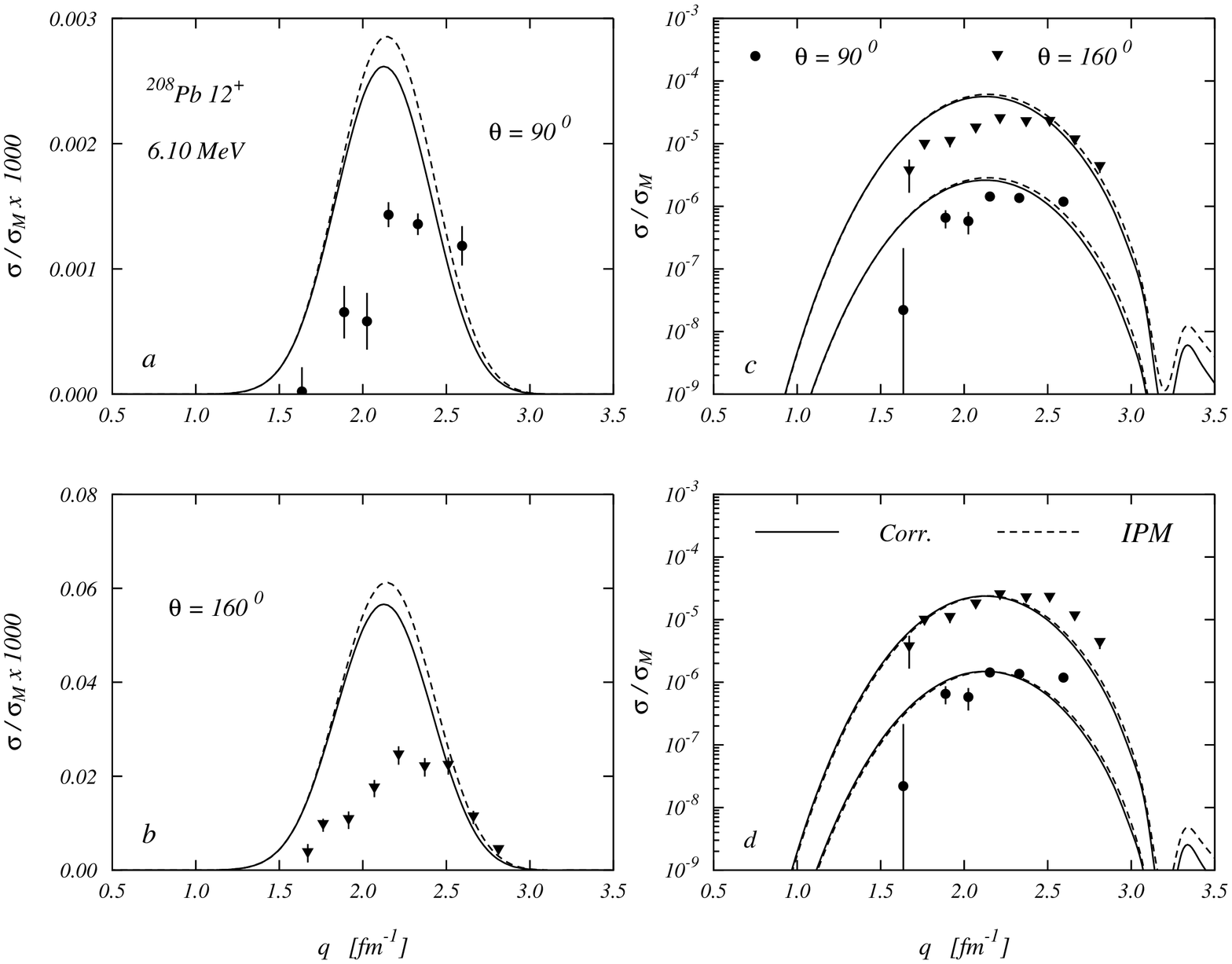}
\caption{\small Inelastic scattering form factors calculated with SRC,
  full lines, and pure MF results, dashed lines.  The left and right
  panels presents the same results in linear and logaritmic scale.
}
\label{fig:pb12}
\end{figure}
The approach briefly outlined is reliable only to describe situations
where nuclear collective effects are negligible. This means that the
dynamics of the excitation should be dominated by single particle
excitations.

For finite nuclear systems, the theory has been applied only for
scalar correlation functions. Specifically we have used the single
particle wave functions and correlations labelled S3 and V8 in ref.
\cite{ari96} and shown in fig. \ref{fig:corr}.

We first applied our model to high angular momentum stretched states
in the discrete spectrum of $^{16}$O, $^{48}$Ca and $^{208}$Pb
\cite{mok00}. Because of the high angular momentum these states
are dominated by a single particle excitation. A typical result of
this study is shown in fig.  \ref{fig:pb12}.  The dashed lines represent
the MF results while the full lines have been obtained by including
the correlation terms of fig. \ref{fig:1p1h}. Clearly the SRC are
unable to explain the quenching required to reproduce the experimental
data.

\begin{figure}
\vspace {-3.0 cm}
\hspace {3.0 cm}
\includegraphics[angle=0,scale=0.6]
       {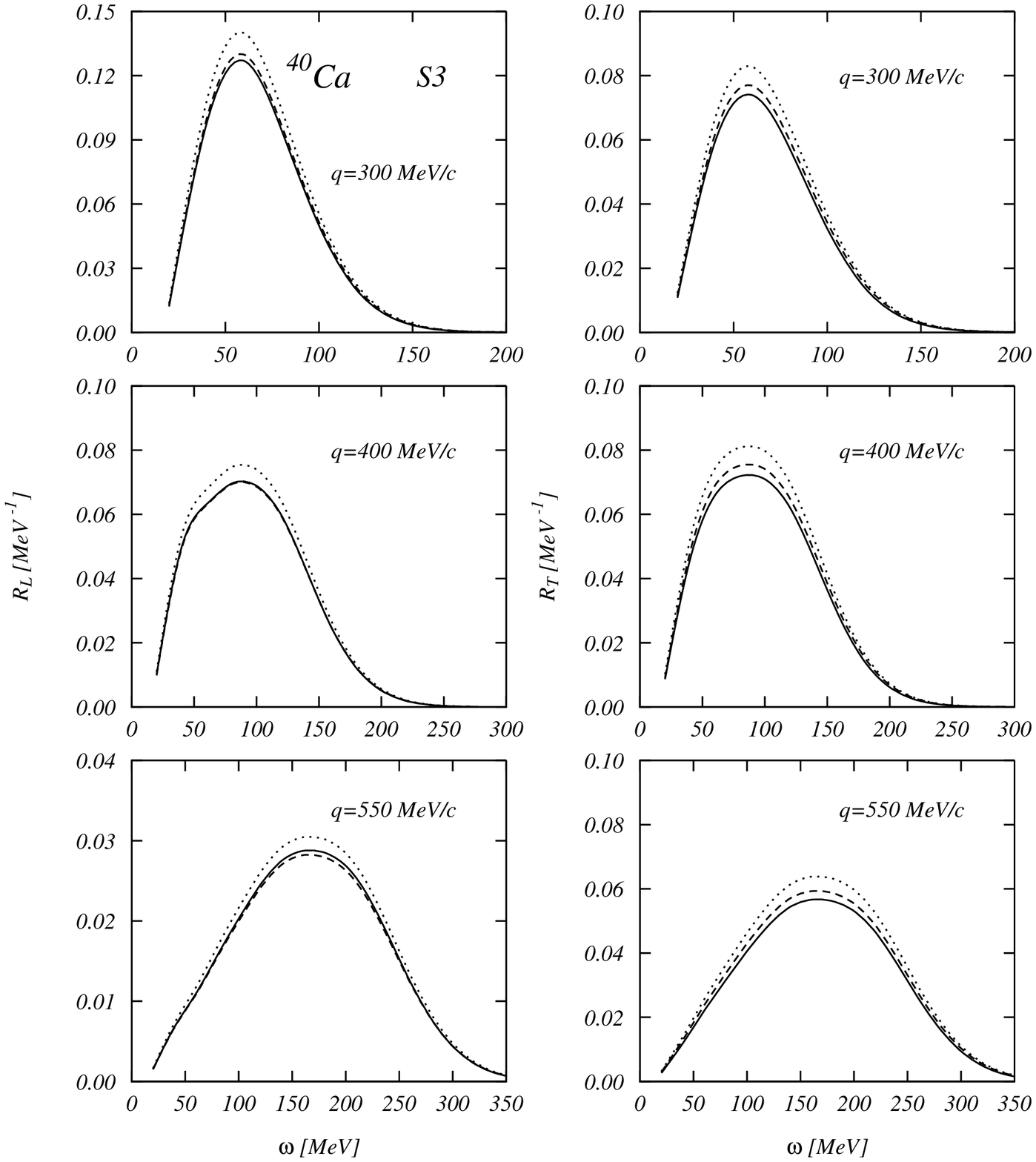}
\caption{\small Charge, R$_L$, and current, R$_T$, inclusive responses. 
Full lines MF responses, dotted lines MF plus 2-point diagrams, dashed
lines MF plus all the diagrams of fig. \protect\ref{fig:1p1h}.
}
\label{fig:resca40}
\end{figure}

We have applied our model in the quasi-elastic region for inclusive
reaction \cite{co01}.  Also in this case the excitation is dominated
by one-particle one-hole excitation dynamics, therefore collective
phenomena are negligible.  However, contrary to the previous case, one
has to consider the re-scattering of the emitted nucleon with the rest
nucleus. This effect, usually named Final State Interaction (FSI),
reduces the maximum values of the responses by a 20-30\% factor
\cite{ama01}.

In fig. \ref{fig:resca40} we show the inclusive electromagnetic
responses calculated in $^{40}$Ca, without the inclusion of the FSI.
The inclusion of only the 2-point diagrams of fig. \ref{fig:1p1h}
slightly increases the MF responses (full lines in fig.
\ref{fig:resca40}). The additional inclusion of the 3-point diagrams
has an opposite effect with respect to that of the 2-point diagrams.
The final difference between the MF and correlated responses is very
small. The effect of the FSI is much larger than that of the SRC
\cite{co01}.

\begin{figure}
\hspace{1.5 cm}
\includegraphics[angle=90,scale=0.5]
       {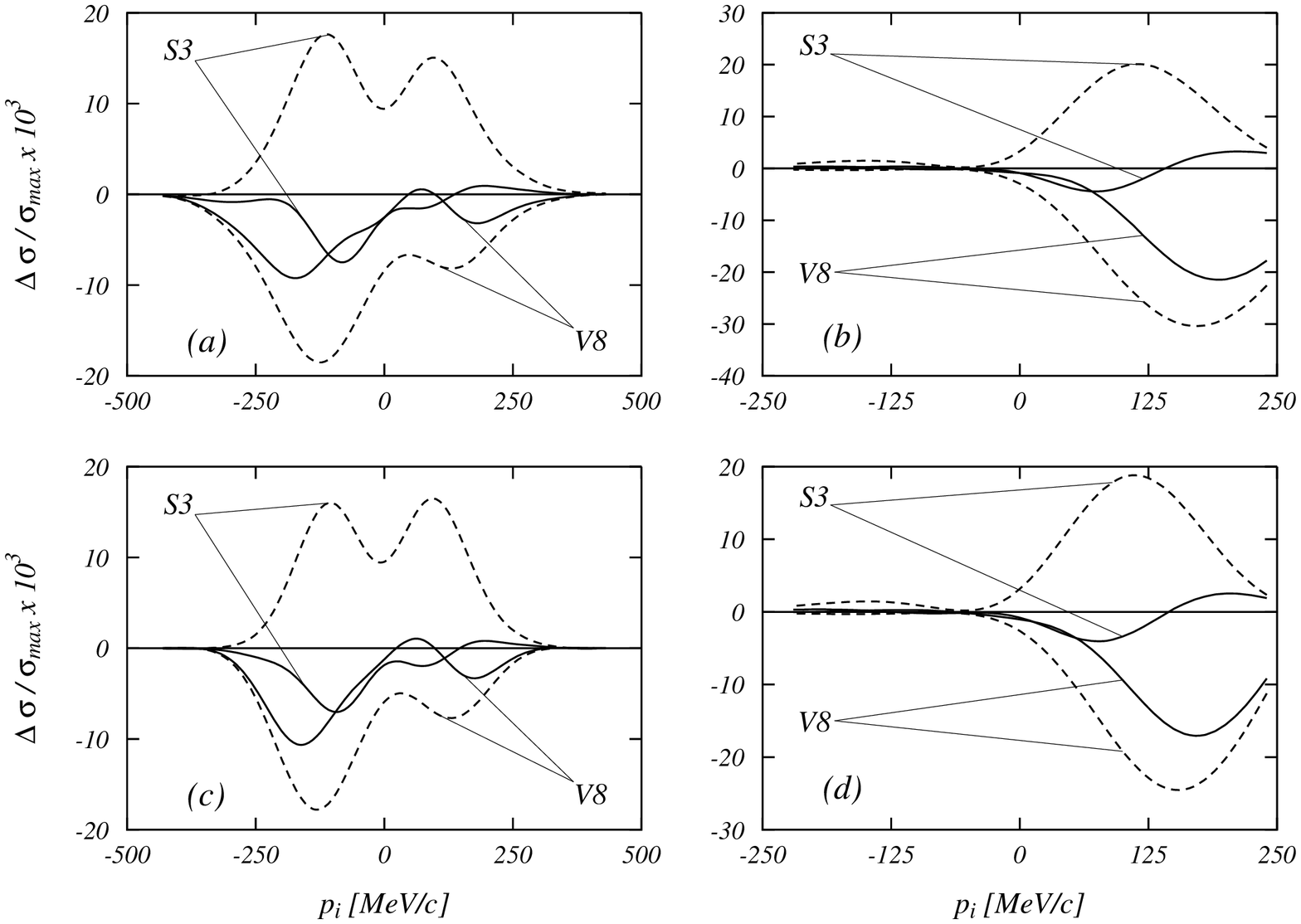}
\caption{\small Normalized differences, eq. \protect\ref{eq:diff}
  ,calculated with the correlation functions of fig.
  \protect\ref{fig:corr}.  The dashed lines have been calculated by
  using two-point diagrams only, while the full lines show the
  inclusion of tree-point diagrams.
}
\label{fig:3point}
%
%
\hspace{1.5 cm}
\includegraphics[angle=90,scale=0.5]
       {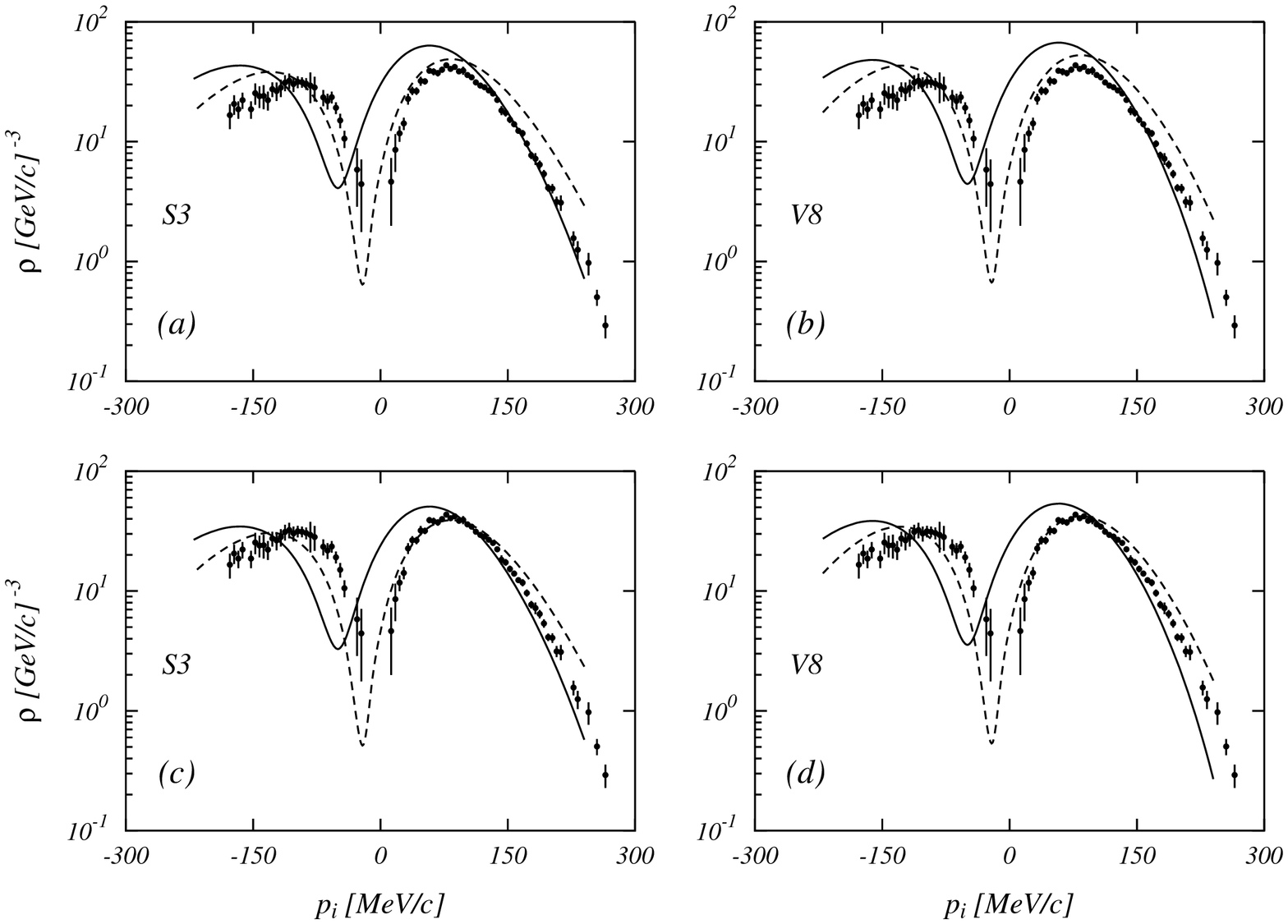}
\caption{\small $^{16}$O (e,e'p)$^{15}$N  
  reduced cross section for the emission of the 1p1/2$^{-1}$ proton
  calculated with the correlation functions of fig.
  \protect\ref{fig:corr}.  The full lines show the results obtained
  with a purely real MF basis, the dashed lines have been obtained
  with an optical potential. The curves in the lower panels have been
  produced by multiplying the upper curves with spectroscopic factor
  of 0.7.
}
\label{fig:nik}
\end{figure}

We have extended our model also to describe (e,e'p) coincidence
experiments on $^{16}$O target \cite{mok01}.  To show the relevance of
the 3-body cluster terms, I show in fig. \ref{fig:3point} the
normalized differences between the one-body response and the
calculation considering two- and three body- SRC terms:

\beq
\Delta=\frac{R^{\rm CORR}(|\bqu|,\omega) - R^{\rm MF}(|\bqu|,\omega)}
               {R^{\rm MF}(|\bqu|,\omega)_{max}}
\label{eq:diff}
\eeq

The results of the two calculations presented in fig. \ref{fig:3point}
have been obtained with the correlations functions of fig.
\ref{fig:corr}. Since the two correlations are rather similar, it is
plausible to expect that the results should not differ very much. This
is not the case when only 2-point diagrams are used, as the dashed
lines of fig. \ref{fig:3point} indicate.  The inclusion of the
three-body clusters, which ensure the correct normalization of the
nuclear final state, produces results which are rather similar for the
two correlation functions.

The relevance of FSI is clear from the comparison with the $^{16}$O
NIKHEF data \cite{leu94} shown in fig. \ref{fig:nik}.  The full lines
have been calculated without FSI and they are distant from the data
points, while the dashed lines considering the FSI via optical
potential are much closer. The SRC are unable to explain the need for
a spectroscopic factor in order to reproduce the data, as the two
lower panels clearly show.

\begin{figure}
\vspace {-1.0 cm}
\hspace {0.75 cm}
\includegraphics[angle=90,scale=0.5]
       {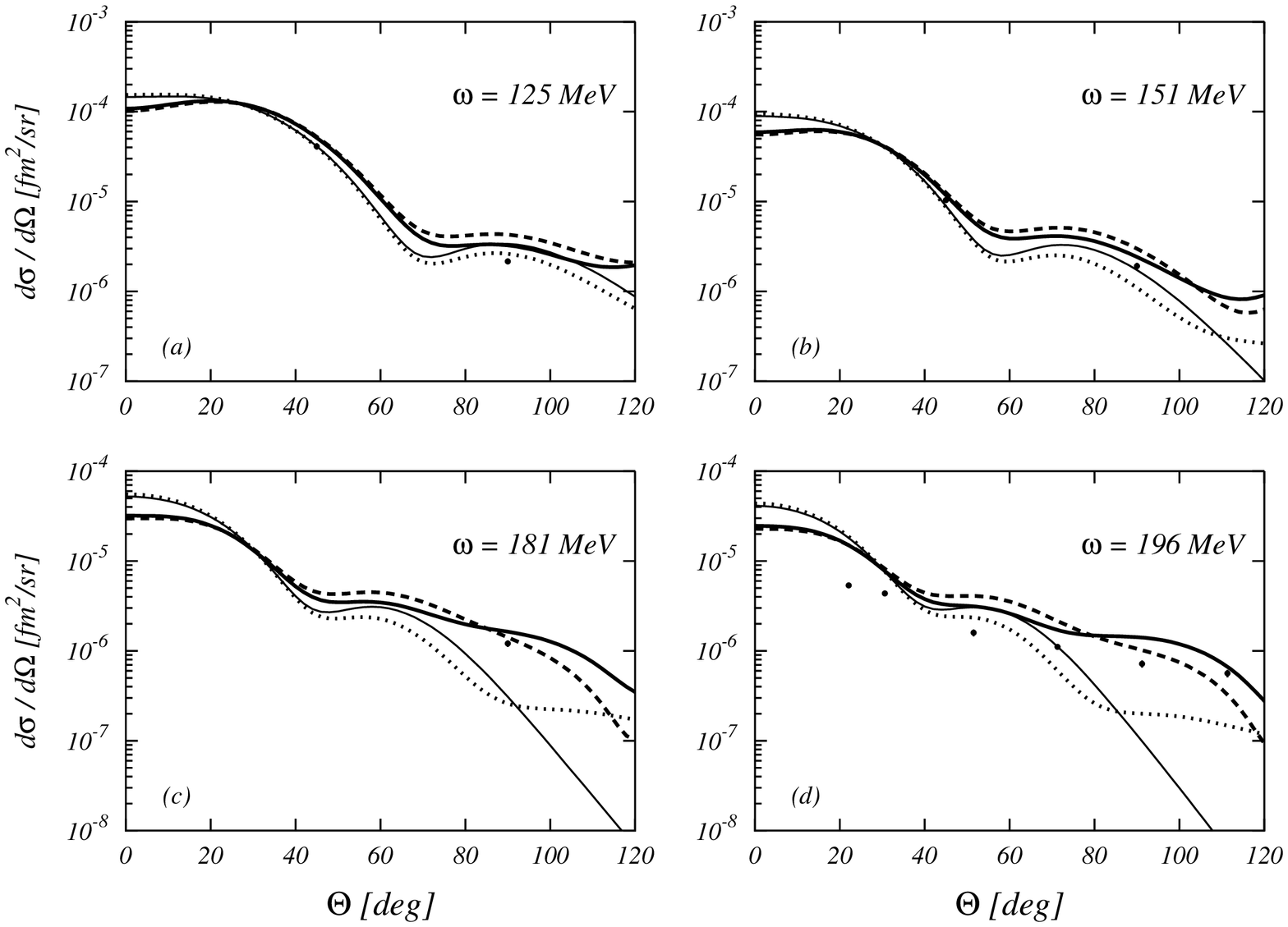}
\vspace{-0.5 cm}
\caption{\small $^{16}$O ($\gamma$,p) cross sections as a function of
  the nucleon emission angle for various values of the photon
  energies.  The thin full lines show the MF results.
  The dotted lines include the effects of the
  S3 correlation. The dashed lines the MEC and the thick full lines
  all the effects.
}
\label{fig:gamma}
%
%
\vspace{-0.5 cm}
\hspace {0.75 cm}
\includegraphics[angle=90,scale=0.5]
       {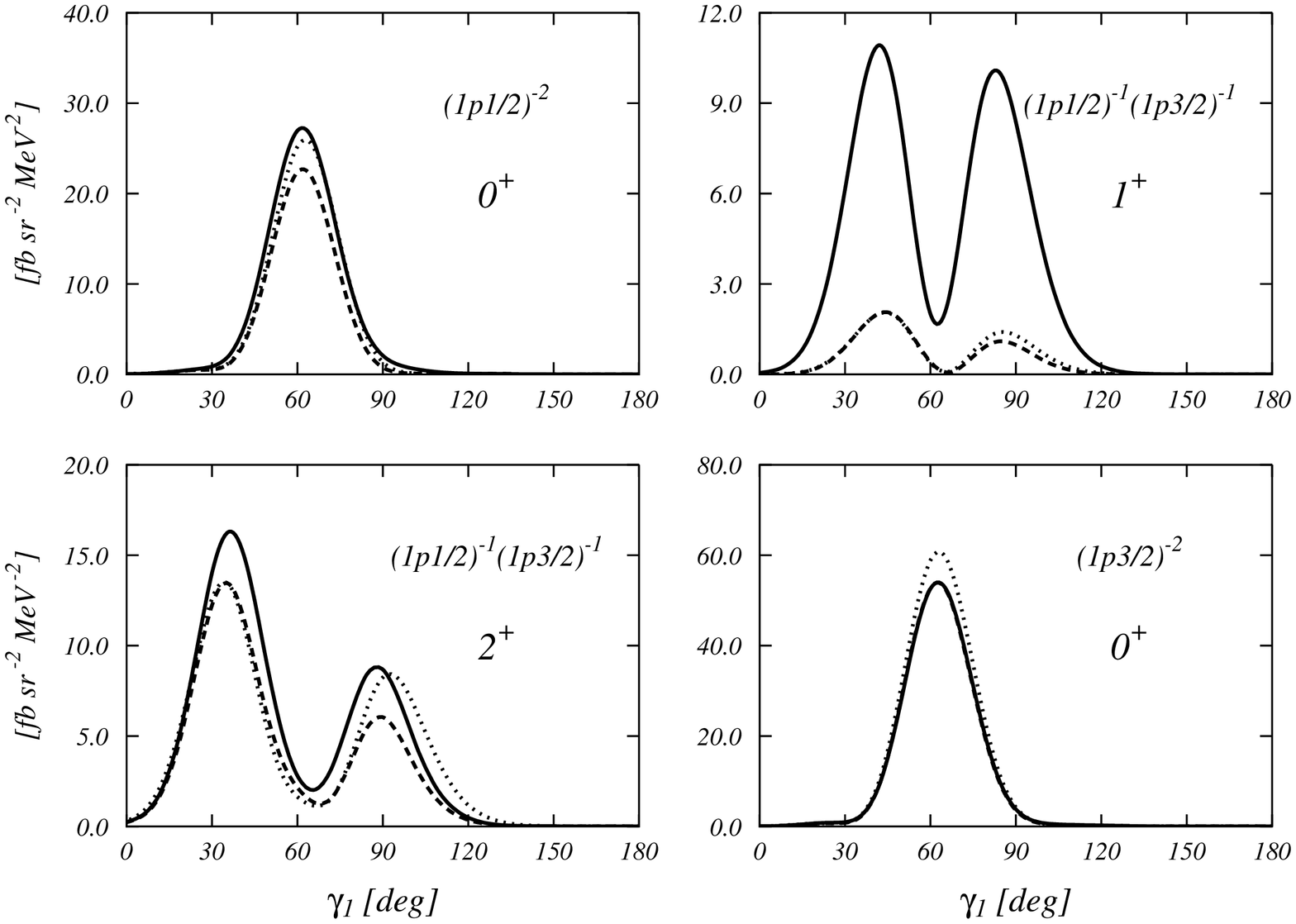}
\vspace{-0.5 cm}
\caption{\small $^{16}$O (e,e'2p) $^{16}$O 
  cross sections as a function of the emission angle of one proton.
  The initial energy of the electron is 500 MeV, the nuclear
  excitation energy 200 MeV, the momentum transfer 300 MeV/c, and the
  second proton is emitted at 60$^0$. The various labels in the
  figures indicates the single particle levels where the two protons
  are emitted and the angular momentum and parity of the residual A-2
  nucleus. The dotted lines have been obtained with 2-point diagrams
  only, the dashed lines by adding the 3-point diagrams and the full
  lines consider also the contribution of the $\Delta$ excitation.
}
\label{fig:ee2p}
\end{figure}

Our model has been also applied to study nucleon emission induced by
real photons. In Fig. \ref{fig:gamma} I show some results of our
investigation.  The thin full lines have been obtained making a pure MF
calculation with one-body current only. The inclusion of SRC,
specifically the S3 of fig. \ref{fig:corr}, produces the dotted lines,
while the inclusion of Meson Exchange Currents (MEC) gives the dashed
lines. The thick full lines show the sum of all the effects
considered.

The effect of the SRC seems to be relevant at high excitation energies
and for large emission angles. Unfortunately also the MEC are important
in this region, and they produce effects as large as those of the SRC,
if not even larger. 

We have applied our model to describe (e,e'2p) processes
\cite{ang02a}. In this case the MF responses induced by a one-body
operator, which in the previous cases were giving the largest
contribution to the cross section, are not present.

In fig. \ref{fig:ee2p} the results for the $^{16}$O (e,e'2p) under the
kinematics explained in the caption are shown.  The dotted lines
represent the results obtained by using only two-point diagrams, the
dashed ones by adding the tree-body diagrams and the full lines
include also the contributions of the excitation of the $\Delta$
resonance.  It is interesting to notice the changes of the relative
importance of the various components of the calculation by changing
the nuclear final state. The common features of all the results shown
in the figure is that the 3-point diagrams reduce the effect of the
2-point ones. This effect of the 3-point diagrams is consistent with
all our previous calculations, and its physical origin is related to
the correct normalization of the nuclear final state, as it has been
detailed discussed in ref. \cite{co01}.

Our calculation show that, even in this case, the contribution of the
two-body currents, consisting here only in the $\Delta$ excitation
term, may cover the SRC effects. The $\Delta$ excitation dominates the
(1p1/2)$^{-2}$ 0$^+$ and the (1p1/2)$^{-1}$ (1p3/2)$^{-1}$ 1$^+$ cross
sections. In the other cases considered its effect is relatively
small.

\section{Conculsions}

Microscopic calculations of nuclear properties require SRC to tame the
strong repulsive core of the realistic nucleon-nucleon interaction. In
our theory the SRC are an input fixed by the minimization of the
hamiltonian expectation value. Within our theory we succeeded in
calculating the properties of $^{16}$O and also of $^{40}$Ca. The
discrepancy with the empirical values of the binding energy is of the
same amount of that of the best nuclear matter calculations. The value
of this quantity shows a scarce sensitivity to the change of the
single particle wave functions, contrary to the density.

To search for effects clearly produced by the SRC we developed a model
describing nuclear responses. Within this model we have investigated
the role of the SRC in (e,e') experiments both in the discrete and in
the continuum spectrum. We have also investigated the single nucleon
emission with both real and virtual photons. These processes are
dominated by the MF transition densities, and the SRC produce very
small effects. In specific situations when they show up, also the
two-body MEC contribute, and the effects of these two processes
strongly mix.

The two-nucleon emission processes are more promising, because in this
case the MF contribution is absent. Also in this case the SRC effects
fight against the MEC ones produced by the $\Delta$ excitation. On the
other hand, the large number of variables could allow for the
possibility of disentangling the two processes.

\vskip 1.cm 
\noindent
{\bf Aknowledgments } \\
It is a pleasure to thank all the colleague and friends whith whom I
have worked on this project: J.E. Amaro, M. Anguiano, F. Arias de
Saavedra, A. Fabrocini, S. Fantoni, P. Folgarait, I.E. Lagaris, A.M.
Lallena, S.R.  Mokhtar.


%
\end{document}